# On the Robust Position Control of Variable Stiffness Series Elastic Actuators

Emre Sariyildiz
School of Mechanical, Materials, Mechatronic and Biomedical Engineering
Faculty of Engineering and Information Sciences, University of Wollongong,
Northfields Avenue Wollongong NSW 2522 Australia
emre@uow.edu.au

*Abstract*— This paper presents robust position control strategies for the novel Variable Stiffness Series Elastic Actuator (VSSEA). By employing a constructed state-space model, two control schemes are developed in a unified framework: a state-feedback controller and a sliding mode controller, both integrated with a second-order disturbance observer (DOb). The proposed framework achieves high-performance motion control by precisely estimating and compensating for internal and external disturbances, while preserving the nominal dynamic response. Simulation results demonstrate that pole-placement-based controllers are highly sensitive to disturbances, whereas Linear Quadratic Regulator (LQR)-based controllers offer improved robustness at the expense of slower dynamics. By incorporating the second-order DOb, robustness is significantly enhanced without degrading time response, and the LQR controller can be tuned solely for performance optimisation. Experimental results confirm that the proposed robust position controllers can be implemented in real world applications. These results highlight the effectiveness of the proposed approach and lay the foundation for future investigations on robust stability and performance under different stiffness settings.

## I. INTRODUCTION

The role of robots is shifting from performing repetitive tasks in structured factory environments to operating in unstructured settings, where they must interact safely and effectively with dynamic environments and humans [1–5]. In these scenarios, safety is a primary requirement, particularly in emerging applications such as collaborative robots (cobots), surgical robots, exoskeletons, and humanoids [5–9]. However, traditional rigid actuators, which dominate industrial robotics, are primarily designed for accuracy and repeatability rather than safe interaction [10]. Their high stiffness and limited compliance can lead to unsafe impacts during contact with humans or uncertain environments, posing serious risks in physical interaction tasks [10 – 12].

To address these limitations, compliant actuation systems have attracted increasing attention over the past decades [13, 14]. By incorporating mechanical compliance into the actuator, robots can achieve safer physical interaction while maintaining task performance. Among these approaches, Series Elastic Actuators (SEAs) have emerged as one of the most widely adopted solutions, enhancing force control accuracy and impact safety while reducing the risk of injury [14 – 17]. Nevertheless, SEA-based actuation has two key drawbacks: (i) constant stiffness may not be suitable for diverse robotic applications, and (ii) the position control problem becomes more challenging due to the introduced compliance [16–21].

To address the first limitation, various types of variable stiffness actuators (VSAs) have been proposed in the literature [22–28]. For instance, antagonistic actuators, which is inspired by human musculoskeletal system, allow adjustable stiffness but suffer from high energy consumption during stiffness modulation [22]. This drawback has been alleviated through novel mechanism designs in different studies [23–28]. Building on this line of research, we recently proposed a novel low-energy-cost VSSEA capable of rapidly modulating stiffness over a wide range [28]. The proposed design enables safe physical interaction in its soft mode while also improving motion control performance across different stiffness levels. However, the introduction of compliance still complicates the position control problem, underscoring the need for advanced robust control strategies that can ensure stability and high performance in the presence of disturbances and uncertainties.

Compared to impedance and force control, relatively few studies have investigated compliant actuators' position control problem [21]. This problem is particularly challenging because the system exhibits fourth order dynamics and the actuator link is highly sensitive to both internal and external disturbances due to mechanical compliance [21]. Conventional approaches for compensating compliance-induced errors (such as Proportional-Integral-Derivative (PID control) [29], singular perturbation control [30], μ-synthesis [31] and intelligent control [32, 33]) have demonstrated limited performance in practical SEA implementations. To address this issue, we recently developed high-performance robust position controllers for SEAs [2, 11]. Despite employing different methods, these controllers could be obtained from the same state-error vector. A unified synthesis framework could therefore provide a clearer understanding of the position control problem in compliant actuation systems.

To this end, this paper proposes a unified synthesis approach for robust position control of actuation systems with intrinsically compliant mechanisms. The traditional state-space dynamic model of compliant actuators is reformulated into a new dynamic representation, enabling the design of robust motion controllers within a unified framework. In the reconstructed model, mismatched disturbances are eliminated, which allows a conventional disturbance observer to be directly applied by feeding back the matched disturbances, with the state-feedback performance controller designed for the reconstructed state vector. Moreover, the new formulation facilitates the direct construction of a sliding manifold based on the redefined error vector. Both the robust state-feedback and sliding mode controllers require disturbance estimate and its first- and second-order derivatives, which are obtained in this study using a second-order DOb. The effectiveness of the proposed robust motion controllers is validated through simulations and experimental results on the novel VSSEA.

The remainder of this paper is structured as follows. Section II introduces the dynamic model of the proposed VSSEA. Section III describes a unified framework for robust position controllers. Section IV reports the simulation and experimental results that validate the proposed approach. Finally, Section V summarises the main conclusions of the study.

## II. Dynamic Model of the Novel Variable Stiffness Actuator

Figure 1 shows the prototype of the novel VSSEA along with its dynamic model. The system is composed of three main subsystems: (i) a rigid actuation unit, consisting of a DC motor and a gearbox, responsible for regulating the equilibrium position; (ii) a variable-stiffness actuation mechanism driven by a direct-drive servo motor; and (iii) the output link. Further details of the novel VSSEA are provided in [28].

Based on the schematic in Fig. 1b, the dynamic model of the VSSEA can be formulated as

*Equilibrium control dynamics:*

$$\begin{aligned} J_l \ddot{\theta}_l + b_l \dot{\theta}_l &= \tau_s - \tau_l^d \\ J_e \ddot{\theta}_e + b_e \dot{\theta}_e &= \tau_e - \tau_{se} - \tau_e^d \end{aligned} \quad (1)$$

where $J_l, b_l, \tau_s$ and $\tau_l^d$ represent the inertia, viscous friction coefficient, spring torque, and disturbances at output link including external load, respectively. In a similar manner, the remaining parameters of the dynamic model can be defined at the motor side using $J_e = J_{me} + N^{-2}J_g$, $b_e = b_{me} + N^{-2}b_g$, $\theta_e = \theta_{me}$, $\tau_e = \tau_{me}$ and $\tau_{se} = N^{-1}\tau_s$, while an equivalent formulation can be obtained at the gear side using $J_e = N^2 J_{me} + J_g$, $b_e = N^2 b_{me} + b_g$, $\theta_e = \theta_g$, $\tau_e = N\tau_{me}$, and $\tau_{se} = \tau_s$. The pseudo disturbance variables $\tau_e^d$ and $\tau_l^d$ can be directly derived from Eq. (1) [34].

*Stiffness modulation dynamics:*

$$J_{ms} \ddot{\theta}_{ms} + b_{ms} \dot{\theta}_{ms} = \tau_{ms} - \tau_s - \tau_{ms}^d \quad (2)$$

where $J_{ms}, b_{ms}, \tau_{ms}, \theta_{ms}$ and $\tau_{ms}^d$ represent the inertia, viscous friction coefficient, motor thrust torque, motor position and pseudo disturbances of the stiffness modulation mechanism, respectively.

Using the Euler-Bernoulli beam theory and small deflection assumption, analytic equations can be derived for the spring torque and the stiffness of the VSSEA as follows:

$$\tau_s(\theta_{ms}, \theta_l) = \frac{\Upsilon_\tau}{\theta_{ms}^3} \sin\left(\frac{\theta_l}{2}\right) \quad (3)$$

$$\Upsilon(\theta_{ms}, \theta_l) = \frac{\Upsilon_k}{\theta_{ms}^3} \cos\left(\frac{\theta_l}{2}\right) \quad (4)$$

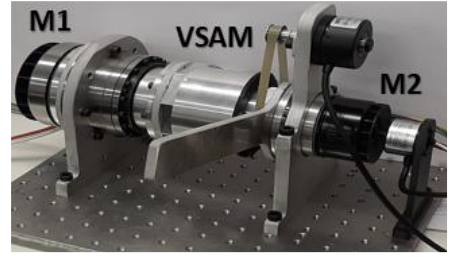

a) Prototype of the variable stiffness actuator.

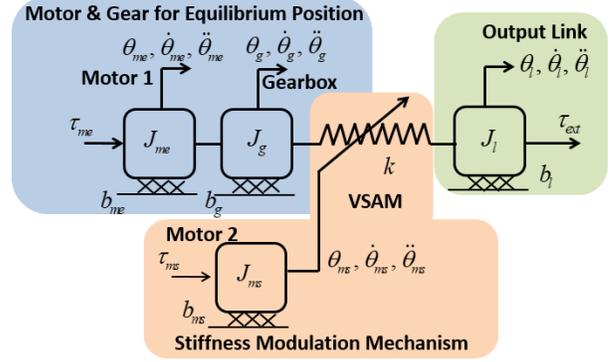

b) Dynamic model of the variable stiffness actuator.

Figure 1: A novel variable stiffness actuator.

where $\Upsilon_\tau$ and $\Upsilon_k$ are the torque and stiffness constant derived in [28].

Equations (3) and (4) show that the stiffness-modulation behaviour is inherently nonlinear, with linear characteristics dominant only within a small deflection range. When synthesising a motion controller, this nonlinear property of the spring must be taken into account to ensure stability and high-performance in practice.

To simplify the control problem, let us obtain a dynamic model for the VSSEA by linearizing the spring at an equilibrium point using $\tau_s = k(\theta_e - \theta_l)$. Using Eq. (1), the state space model of the actuator is derived as follows:

$$\dot{\mathbf{x}}(t) = \mathbf{A}\mathbf{x}(t) + \mathbf{B}u(t) - \mathbf{D}(t) \quad (5)$$

where $\mathbf{A} = \begin{bmatrix} 0 & 1 & 0 & 0 \\ -k/J_l & -b_l/J_l & k/J_l & 0 \\ 0 & 0 & 0 & 1 \\ k/J_e & 0 & -k/J_e & -b_e/J_e \end{bmatrix}$, $\mathbf{B} = \begin{bmatrix} 0 \\ 0 \\ 0 \\ 1/J_e \end{bmatrix}$, $\mathbf{D} = \begin{bmatrix} 0 \\ \tau_l^d/J_l \\ 0 \\ \tau_e^d/J_e \end{bmatrix}$,

$\mathbf{x} = \begin{bmatrix} \theta_l & \dot{\theta}_l & \theta_e & \dot{\theta}_e \end{bmatrix}^T$, and $u = \tau_e$.

In Eq. (5), $\tau_e^d/J_e$ and $\tau_l^d/J_l$ denote the matched and mismatched disturbances acting on the actuator dynamics. Because matched disturbances enter through the same channel as the control input $u(t)$ (i.e., through the fourth channel of the input matrix **B**), they can be directly compensated by feeding back their estimates. By contrast, conventional DOb-based robust control cannot eliminate mismatched disturbances, which do not align with the input channel and can at best be attenuated indirectly [34, 35]. Despite its wide adoption, the dynamic model given in Eq. (5) is not useful in conventional DOb-based robust motion controller synthesis.

### III. ROBUST POSITION CONTROL OF THE NOVEL VARIABLE STIFFNESS ACTUATOR

*Output Link Position Control:*

To synthesize robust position controllers, let us construct a more useful dynamic model for the VSSEA. The state-space dynamic model given in Eq. (5) can be rewritten as follows:

$$\dot{\mathbf{x}}(t) = \mathbf{\Gamma}\mathbf{x}(t) + \mathbf{B}u(t) - \mathbf{\Pi}(t) \quad (6)$$

where $\mathbf{\Gamma} = \begin{bmatrix} 0 & 1 & 0 & 0 \\ 0 & 0 & 1 & 0 \\ 0 & 0 & 0 & 1 \\ 0 & 0 & 0 & 0 \end{bmatrix}, \mathbf{\Pi} = \begin{bmatrix} 0 \\ \Pi_2 \\ 0 \\ \Pi_4 \end{bmatrix}$ in which the disturbances are

$\Pi_2 = \frac{1}{J_l}\left(J_l\theta_e + k(\theta_l - \theta_e) + b_l\dot{\theta}_l + \tau_l^d\right)$ and $\Pi_4 = \frac{1}{J_e}\left(k(\theta_e - \theta_l) + b_e\dot{\theta}_e + \tau_e^d\right)$.

Equation (6) indicates that the output link is governed indirectly: the controller sets the equilibrium motor position, and the spring transmits the resulting torque to the link. Therefore, we must derive a control law that maps the desired link reference to the corresponding equilibrium-motor command using the dynamics in (6).

Let us define the trajectory tracking error using $e_l = \theta_l^{ref} - \theta_l$ where $\theta_l^{ref}$ represents the position reference of the output link. From Eq. (6), the following error dynamics can be obtained for the position control problem of the VSSEA.

$$\dot{\mathbf{e}}(t) = \mathbf{\Gamma}\mathbf{e}(t) - \mathbf{B}u(t) + \tilde{\mathbf{\Pi}}(t) \quad (7)$$

where $\tilde{\mathbf{\Pi}} = \begin{bmatrix} 0 & 0 & 0 & \tilde{\Pi}_4 \end{bmatrix}^T$ is the matched disturbance vector in which $\tilde{\Pi}_4 = \ddot{\Pi}_2 + \Pi_4$, and $\mathbf{e} = \begin{bmatrix} x_1^{ref} - x_1 \\ x_2^{ref} - x_2 \\ x_3^{ref} - x_3 + \Pi_2 \\ x_4^{ref} - x_4 + \dot{\Pi}_2 \end{bmatrix} = \begin{bmatrix} \theta_l^{ref} - \theta_l \\ \dot{\theta}_l^{ref} - \dot{\theta}_l \\ \ddot{\theta}_l^{ref} - \theta_e + \Pi_2 \\ \dddot{\theta}_l^{ref} - \dot{\theta}_e + \dot{\Pi}_2 \end{bmatrix}$.

The matched disturbance $(\tilde{\Pi}_4)$ acting on the error dynamics in the fourth channel can derived by taking the derivative of the last term in the position control error vector **e**.

As shown in Eq. (7), the position-error dynamics contain only matched disturbances. This is achieved by a suitable state-space reparameterisation that channels all disturbance terms through the input matrix. Consequently, a conventional DOb-based robust controller can be designed by directly using Eq. (7). However, constructing Eq. (7) requires estimating the disturbance and its first-order time derivative, which generally increases the noise sensitivity of robust motion control systems in practice. This is a common trade-off in DOb-based robust position control of compliant robotic systems [34]. The proposed dynamic model, nevertheless, provides a flexible foundation for implementing various robust motion control strategies.

*Robust State Feedback Controller:*

Let us begin with a robust state-feedback control scheme using DOb. The controllability of the dynamic model follows from

$$\text{rank}\left(\begin{bmatrix} \mathbf{B} & \mathbf{\Gamma}\mathbf{B} & \mathbf{\Gamma}^2\mathbf{B} & \mathbf{\Gamma}^3\mathbf{B} \end{bmatrix}\right) = 4 \quad (8)$$

hence the four-state model is controllable and we can freely adjust the error dynamics of the position controller. The robust position control signal can be generated using

$$u(t) = J_e\left(\mathbf{K}^T\hat{\mathbf{e}}(t) - \hat{\tilde{\Pi}}_4(t)\right) \quad (9)$$

where $\mathbf{K} = \begin{bmatrix} k_1 & k_2 & k_3 & k_4 \end{bmatrix}^T$ is the state feedback control gain that can be tuned using different state space control techniques such as pole placement, $\hat{\tilde{\Pi}}_4(t) = \hat{\dot{\Pi}}_2 + \hat{\Pi}_4$ is the estimated matched disturbance given in Eq. (7) and $\hat{\mathbf{e}}(t)$ is the estimation of the constructed error state vector given by.

$$\hat{\mathbf{e}} = \begin{bmatrix} \theta_l^{ref} - \theta_l \\ \dot{\theta}_l^{ref} - \dot{\theta}_l \\ \ddot{\theta}_l^{ref} - \theta_e + \hat{\Pi}_2 \\ \dddot{\theta}_l^{ref} - \dot{\theta}_e + \hat{\dot{\Pi}}_2 \end{bmatrix} \quad (10)$$

As indicated by Eqs. (9) and (10), synthesising the proposed robust motion controller requires estimates of the disturbance and its first- and second-order time derivatives. A higher-order DOb synthesis is outlined in the next section.

The stability of the proposed robust state feedback controller can be proved using the following Lyapunov function candidate.

$$V = \mathbf{e}^T\mathbf{P}\mathbf{e} \quad (11)$$

where **P** is a positive definite symmetric matrix that satisfies

$$\left(\mathbf{\Gamma} - \mathbf{B}\mathbf{K}^T\right)^T\mathbf{P} + \mathbf{P}\left(\mathbf{\Gamma} - \mathbf{B}\mathbf{K}^T\right) = -\mathbf{Q} \quad (12)$$

where **Q** is also a positive definite symmetric matrix. Since the dynamic model is controllable, we can select **Q** and **P** matrices that meet this requirement.

The derivative of Eq. (11) is as follows:

$$\dot{V} = -\mathbf{e}^T\mathbf{Q}\mathbf{e} + 2\mathbf{e}^T\mathbf{P}\left(\tilde{\Pi}_4 - \hat{\tilde{\Pi}}_4\right) \\ \leq -\lambda_{\min}(\mathbf{Q})\|\mathbf{e}\|^2 + 2\|\mathbf{e}\|\lambda_{\max}(\mathbf{P})\left\|\tilde{\Pi}_4 - \hat{\tilde{\Pi}}_4\right\| \quad (13)$$

Equation (13) shows that the proposed robust state feedback controller is uniformly ultimately bounded when the disturbance estimation error is bounded. Also, asymptotic stability is achieved when disturbances are precisely cancelled by the DOb.

*Sliding Mode Robust Position Controller:*

Let us now synthesise a sliding mode control based robust position controller using the dynamic model in Eq. (7). A scalar sliding variable can be defined as follows:

$$\sigma(t) = \mathbf{S}^T \mathbf{e}(t) \tag{14}$$

where $\mathbf{S} = \begin{bmatrix} s_1 & s_2 & s_3 & 1 \end{bmatrix}^T$ and the coefficients $s_1$, $s_2$ and $s_3$ are free design parameters that shape the dynamics on the sliding manifold $\sigma(t)$ and $\mathbf{e}(t)$ is given in Eq. (7). The derivative of Eq. (14) is as follows:

$$\dot{\sigma}(t) = \mathbf{S}^T \dot{\mathbf{e}}(t) = -\frac{1}{J_e}\tau_e + \left( \ddot{\theta}_l^{ref} + \ddot{\Pi}_2 + \Pi_4 + \sum_{i=1}^{3} s_i e_{i+1} \right) \tag{15}$$

where $e_2 = \dot{\theta}_l^{ref} - \dot{\theta}_l, e_3 = \ddot{\theta}_l^{ref} - \ddot{\theta}_e + \Pi_2, \dot{e}_3 = \dddot{\theta}_l^{ref} - \dot{\theta}_e + \dot{\Pi}_2$ are error states of the dynamic model in Eq. (7).

The robust position control signal is generated using

$$\tau_e = J_e \rho \, \text{sgn}(\hat{\sigma}) + J_e \hat{\delta} \tag{16}$$

where $\hat{\delta} = \ddot{\theta}_l^{ref} + \hat{\dot{\Pi}}_2 + \hat{\Pi}_4 + \sum_{i=1}^{3} s_i \hat{e}_{i+1}$ and $\hat{\bullet}$ represents the estimated $\bullet$.

The stability of the proposed robust controller can be proved using the following Lyapunov function candidate.

$$V = \frac{1}{2}\sigma^2 \tag{17}$$

The derivative of Eq. (17) is

$$\dot{V} = \sigma\left(-\rho \, \text{sgn}(\sigma) - (\delta - \hat{\delta})\right) \leq -\left(\rho - |\delta - \hat{\delta}|\right)|\sigma| \tag{18}$$

When the disturbance estimation error is bounded, a bounded position control error is achieved. Similarly, the robust position controller is asymptotically stable when the DOb precisely cancels disturbances.

*Stiffness Modulation Control:*

To modulate the stiffness of the VSSEA, the stiffness-modulation motor must track its reference precisely while rejecting disturbances arising from beam deflection and modelling uncertainties. This is achieved by regulating the motor position through a conventional DOb-based robust control loop.

$$\tau_{ms} = K_d\left(\dot{\theta}_{ms}^{ref} - \dot{\theta}_{ms}\right) + K_p\left(\theta_{ms}^{ref} - \theta_{ms}\right) + \hat{\tau}_{ms}^d \tag{19}$$

where $K_p$ and $K_d$ are the proportional and derivative gains of the performance controller and $\hat{\tau}_{ms}^d$ is the estimated disturbances acting on the stiffness modulation mechanism.

## IV. SECOND-ORDER DOB

To implement the proposed robust position controllers, let us design a second order DOb that can estimate disturbances and their first and second order derivatives [35, 36].

Lest us consider the following auxiliary variable vectors.

$$\begin{aligned} \mathbf{a}_0(t) &= \mathbf{D}(t) + g_0 \mathbf{x} \\ \mathbf{a}_1(t) &= \dot{\mathbf{D}}(t) + g_1 \mathbf{x} \\ \mathbf{a}_2(t) &= \ddot{\mathbf{D}}(t) + g_2 \mathbf{x} \end{aligned} \tag{20}$$

where $\mathbf{a}_0$, $\mathbf{a}_1$ and $\mathbf{a}_2$ are auxiliary variables constructed from the disturbance vector and its first- and second-order derivatives, together with the system state multiplied by the observer gains $g_0$, $g_1$ and $g_2$.

The derivative of Eq. (20) is as follows:

$$\dot{\mathbf{a}}(t) = \Lambda_a \mathbf{a}(t) + \Lambda_u u(t) + \Lambda_x \mathbf{x}(t) + \Lambda_D \tag{21}$$

where $\Lambda_a = \begin{bmatrix} -g_0 \mathbf{I}_4 & \mathbf{I}_4 & \mathbf{0}_4 \\ -g_1 \mathbf{I}_4 & \mathbf{0}_4 & \mathbf{I}_4 \\ -g_2 \mathbf{I}_4 & \mathbf{0}_4 & \mathbf{0}_4 \end{bmatrix}, \Lambda_u = \begin{bmatrix} g_0 \mathbf{B} \\ g_1 \mathbf{B} \\ g_2 \mathbf{B} \end{bmatrix}, \Lambda_x = \begin{bmatrix} g_0(\mathbf{A} + g_0 \mathbf{I}) - g_1 \mathbf{I} \\ g_1(\mathbf{A} + g_0 \mathbf{I}) - g_2 \mathbf{I} \\ g_2(\mathbf{A} + g_0 \mathbf{I}) \end{bmatrix}$

and $\Lambda_D = \begin{bmatrix} \mathbf{0}_4 & \mathbf{0}_4 & \dddot{\mathbf{D}}(t) \end{bmatrix}^T$.

An observer that can estimate the disturbances and their first and second order derivatives can be designed by neglecting the third order derivative of disturbances in Eq. (21) as follows:

$$\dot{\hat{\mathbf{a}}}(t) = \Lambda_a \hat{\mathbf{a}}(t) + \Lambda_u u(t) + \Lambda_x \mathbf{x}(t) \tag{22}$$

where $\hat{\mathbf{a}}(t)$ represents auxiliary variable vector estimate.

Subtracting Eq. (22) from Eq. (21) yields

$$\dot{\mathbf{e}}_a(t) = \Lambda_a \mathbf{e}_a(t) + \Lambda_D \tag{23}$$

where $\mathbf{e}_a(t) = \mathbf{a}(t) - \hat{\mathbf{a}}(t)$ represents the estimation error.

When the observer gains are chosen so that the error-dynamics matrix $\Lambda_a$ is Hurwitz (all eigenvalues strictly in the open left half-plane), the disturbance-estimation error $\mathbf{e}_a(t)$ is exponentially stable and satisfies

$$\|\mathbf{e}_a(t)\| \leq \exp(-\lambda_{\min} t)\|\mathbf{e}_a(t_0)\| + \lambda_{\min}^{-1} |\Lambda_D| \tag{24}$$

where $\lambda_{\min}$ represents minimum eigenvalue of $\Lambda_a$.

## V. SIMULATIONS AND EXPERIMENTS

This section verifies the proposed robust position controllers via simulations and experiments. The output link position is controlled using Maxon EC90 flat motor and a 1:100 ratio harmonic drive, while the stiffness modulation was controlled using a Maxon EC60 direct drive motor. Both motors are controlled using Escon motor drivers and a Simulink-based real-time motion control system with the sampling rate 1ms.

Let us begin with the position control simulations of the VSSEA. Figure 2 presents both regulation and trajectory tracking results when a state-feedback controller is designed using pole-placement and Linear Quadratic Regulator (LQR) techniques. Nonlinear, time-varying external disturbances are

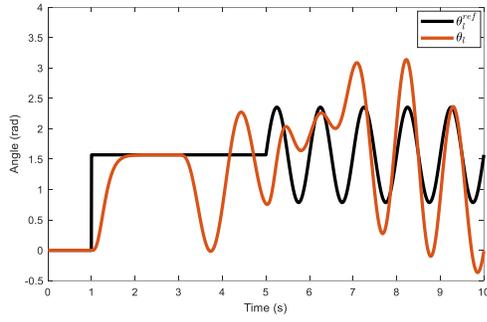
a) Regulation and trajectory tracking control using pole-placement.

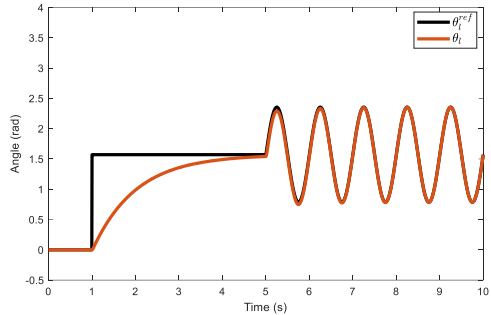
b) Regulation and trajectory tracking control using LQR.

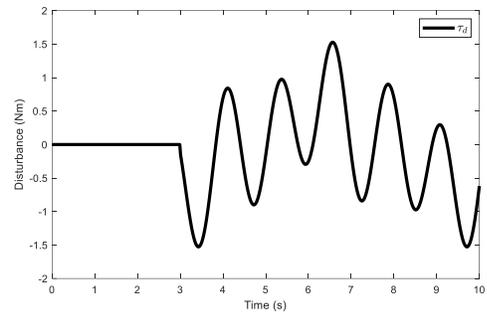
c) Disturbances.
Figure 2: Position control using a state feedback controller.

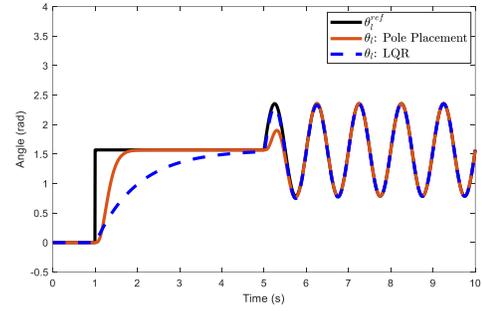
a) Robust state feedback controller when the performance controller is tuned using pole placement and LQR.

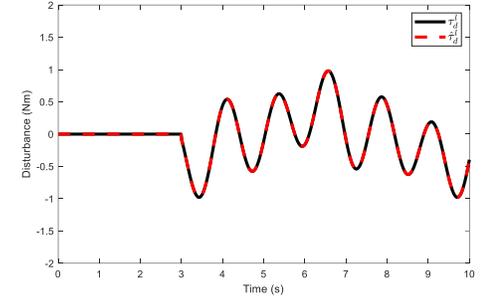
b) Disturbances and their estimations.

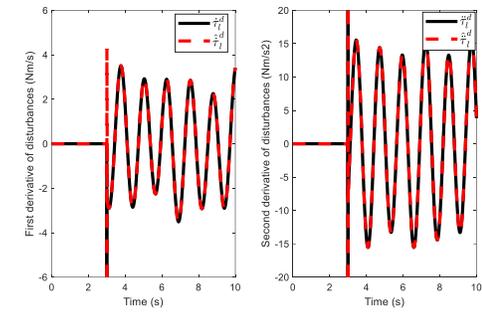
c) First and second derivatives of disturbances and their estimates.
Figure 3: Position control simulations using robust motion controller.

introduced after 3 seconds. As shown in Fig. 2a, the pole-placement-based controller is highly sensitive to disturbances, leading to a significant increase in position tracking error. In contrast, the LQR controller improves robustness against disturbances; however, this comes at the cost of a slower system response. This trade-off highlights the inherent limitation of relying solely on LQR for robustness enhancement.

To enhance the performance of the position control system, let us integrate a second-order DOb with the feedback controller, thereby synthesising the proposed robust position control scheme. The corresponding simulation results are illustrated in Figure 3. As shown in Fig. 3a, the proposed robust state-feedback controller markedly improves disturbance rejection while maintaining the original time response of the system, i.e., it does not slow down the dynamics. Since robustness can be independently enhanced by the DOb, the proposed controller can also be applied in conjunction with LQR, as depicted in the figure. In this case, the LQR controller can be tuned solely for performance optimisation, while the DOb compensates for external disturbances.

Implementation of the proposed robust state-feedback controller requires estimating the disturbances together with their first- and second-order derivatives. Figures 3b and 3c demonstrate that the high-order DOb is capable of providing accurate estimates of disturbances and their higher-order derivatives, thereby enabling the synthesis of the proposed control strategy.

Figure 4 demonstrates that both the proposed robust state-feedback controller and the sliding mode controller achieve high-performance robust position control in real-world implementation. The VSSEA is able to accurately follow the reference trajectory even when subjected to disturbances applied between 3 and 10 seconds. These experimental results, together with the simulation studies, provide strong verification of the effectiveness of the proposed control strategies.

## VI. CONCLUSIONS

This paper has proposed robust position control schemes for the novel VSSEA based on state-feedback and sliding mode control. The controllers achieve high-performance motion control by effectively suppressing both internal and external disturbances through the integration of a second-order DOb and a state-space controller synthesis method. The effectiveness of

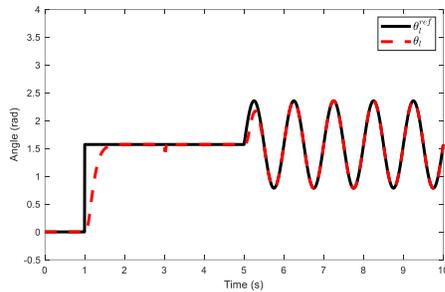

a) Position control using robust state feedback controller.

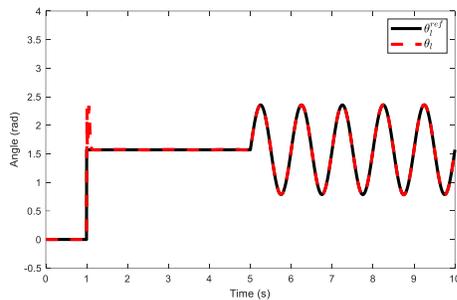

b) Position control using sliding mode controller.

Figure 4: Position control experiments using robust motion controller

the proposed robust controllers has been validated through simulations and experimental results for both regulation and trajectory tracking tasks. Future work will focus on extending the analysis to varying stiffness levels, as well as conducting a more detailed investigation of robust stability and performance under broader operating conditions.